\documentclass{emulateapj}

\usepackage{graphicx}
\usepackage{float}
\usepackage{amsmath}
\usepackage{epsfig,floatflt}

\begin{document}

\title{The effect of asymmetric beams in the Wilkinson Microwave Anisotropy Probe experiment }

\author{I. K. Wehus\altaffilmark{1}, L. Ackerman\altaffilmark{2},
  H.\ K.\ Eriksen\altaffilmark{3,4}, and N. E. Groeneboom\altaffilmark{3,4}}

\email{i.k.wehus@fys.uio.no}
\email{lotty@caltech.edu}
\email{h.k.k.eriksen@astro.uio.no}
\email{leuat@irio.co.uk}

\altaffiltext{1}{Department of Physics, University of
  Oslo, P.O.\ Box 1048 Blindern, N-0316 Oslo, Norway}

\altaffiltext{2}{California Institute of Technology, Pasadena, CA
  91125}

\altaffiltext{3}{Institute of Theoretical Astrophysics, University of
  Oslo, P.O.\ Box 1029 Blindern, N-0315 Oslo, Norway}

\altaffiltext{4}{Centre of Mathematics for Applications, University of
  Oslo, P.O.\ Box 1053 Blindern, N-0316 Oslo, Norway}


\begin{abstract}
We generate simulations of the CMB temperature field as observed by
the WMAP satellite, taking into account the detailed shape of the
asymmetric beams and scanning strategy of the experiment, and use
these to re-estimate the WMAP beam transfer functions. This method
avoids the need of artificially symmetrizing the beams, as done in the
baseline WMAP approach, and instead measures the total convolution
effect by direct simulation. We find only small differences with
respect to the nominal transfer functions, typically less than 1\%
everywhere, and less than 0.5\% at $\ell < 400$. The net effect on the
CMB power spectrum is less than 0.6\%. The effect on all considered
cosmological parameters are negligible. For instance, we find that the
spectral index of scalar perturbations after taking into account the
beam asymmetries is $n_{\textrm{s}} = 0.964\pm0.014$, corresponding to
a negative shift of $-0.1\sigma$ compared to the previously released
WMAP results. Our CMB sky simulations are made publicly available, and
can be used for general studies of asymmetric beam effects in the WMAP
data.
\end{abstract}

\section{Introduction}
\label{sec:introduction}

Without doubt, the angular CMB power spectrum is today our single most
important source of cosmological information. Perhaps the most
striking demonstration of this fact to date is the WMAP experiment,
\citep{bennett:2003, hinshaw:2007, hinshaw:2009} which has allowed
cosmologists to put unprecedented constraints on all main cosmological
parameters, as well as ruling out vast regions of the possible model
spaces. Similarly, in only a few years from now Planck will finally
provide the definitive measurements of the temperature power spectrum,
as well as polarization spectra with unprecedented accuracy. This will
certainly lead to similar advances in our knowledge about the history
of our universe.

Each of these experiments observes  the CMB field by scanning the sky
with an instrumental beam of finite resolution. This operation
effectively corresponds to averaging over beam-sized angular scales,
and is expressed technically either in pixel space by a convolution of
the beam with the underlying sky, or in harmonic space by a
multiplication of the two corresponding sets of harmonic expansion
coefficients. For simplicity, the harmonic space expansion of the beam
is typically expressed in terms of Legendre coefficients of an
(azimuthally symmetric) effective beam response. This function is
often called ``the beam transfer function'', $b_{\ell}$.

Before it is possible to make unbiased cosmological inferences based
on the CMB power spectrum, it is of critical importance to know the
beam transfer function to high precision, as an error in the beam
function translates into a direct bias in the estimated power
spectrum. This in turn requires detailed knowledge about the beam
response function on the sky for each experiment. For a full
description of the WMAP beam estimation process and final model, see
\citet{page:2003}, \citet{jarosik:2007} and \citet{hill:2009}.

The impact of asymmetric beams may also be important for applications
other than power spectrum estimation. One example of special interest
to us is the assessment of non-Gaussianity and violation of
statistical isotropy. Specifically, \citet{ackerman:2007} considered a
model based on violation of rotational invariance in the early
universe, and derived explicit parametric expressions for the
corresponding observational signature. Then, in a follow-up paper
\citet{groeneboom:2009a} analysed the 5-year WMAP data with respect to
this model and, most surprisingly, found a detection at the
$3.8\sigma$ confidence level. Given that this was a most unexpected
result, several questions concerning systematic errors in the WMAP
data were considered, in particular those due to residual foregrounds,
correlated noise and asymmetric beams. However, it was shown in the
same paper that neither foregrounds nor correlated noise were viable
explanations, while the question of asymmetric beams was left
unanswered, due to lack of proper simulation machinery. This question
provided our initial motivation for considering the problems studied
in this paper.

The starting point for tackling the asymmetric beam problem for WMAP
is a set of beam maps released by the WMAP team, two for each
differencing assembly (DA), denoted A and B, respectively. These maps
were derived by observing Jupiter for extended periods of time. Then,
in order to derive the proper beam transfer functions, the WMAP team
adopted a computationally fast and convenient approach: They first
symmetrized the effective beam for each DA, collapsing the information
in the A and B sides into one common function, and then computed the
Legendre transform of the corresponding radial profile. However, for
this to be an accurate approximation, one must on the one hand assume
that the beams on the two sides are very similar, and on the other
hand either assume that both beams are intrinsically circularly
symmetric, or that all pixels on the sky are observed from all angles
an equal number of times due to the scanning strategy. In reality none
of these conditions are met, and one may therefore ask whether there
might be any residual effect due to the combination of an asymmetric
beam and anisotropic scanning in the WMAP beam functions.

This problem was addressed analytically by \citet{hinshaw:2007}, who
derived an approximate expression for the expected power spectrum bias
due to asymmetric beams in the WMAP data. Their conclusion was that
such effects were $\lesssim$1\% for the 3-year WMAP data. 

In this paper, we revisit the question of asymmetric beams in WMAP
with two main goals. First, we seek to estimate the effective beam
transfer functions for each WMAP DA, taking into account the full
details of the asymmetric beams and specifics of the WMAP scanning
strategy by direct simulation. This way, we check whether the analytic
approximations presented by \citet{hinshaw:2007} are valid. Second, we
want to produce a set of high-fidelity simulated CMB sky maps, with
beam properties as close as possible to those observed by WMAP, that
can later be used for general studies of asymmetric beam effects in
WMAP.

\section{Pipeline overview}
\label{sec:pipeline}

In this section we summarize the methods and algorithms used in this
paper. Note that none of the individual steps described below are
original to this paper, and only the main ideas will therefore be
discussed in the following.

\begin{figure}[t]
\mbox{\epsfig{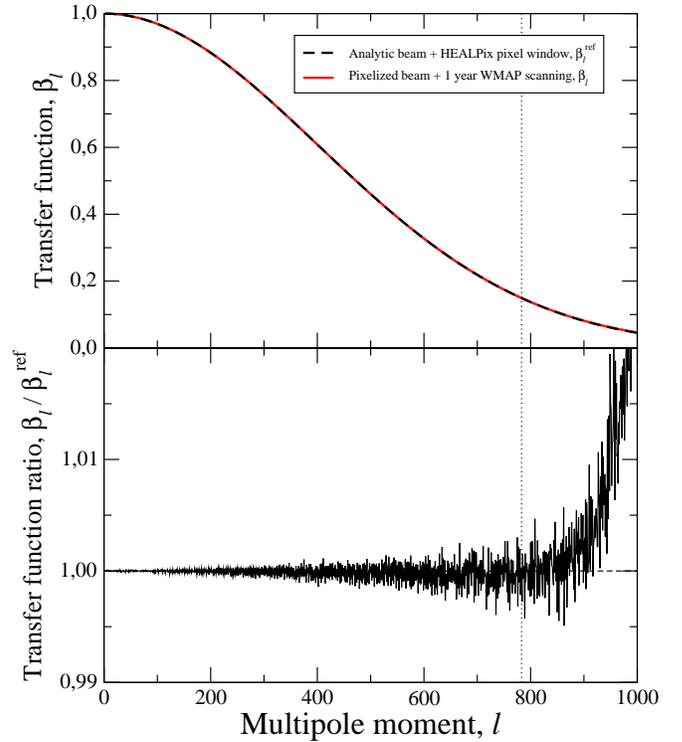}}
\caption{Top panel: Comparison between the transfer functions,
  $\beta_{\ell}$, for a Gaussian beam of $20'$ FWHM. This was computed
  from a pixelized beam map and with the WMAP V1 scanning strategy
  (red line), and alternatively, by using the well-known analytic expression for the
  Legendre transform of a Gaussian beam (Equation
  \ref{eq:gaussian_beam}) and isotropized HEALPix pixel window (black
  dashed line). The vertical dotted line indicates the multipole
  moment, $\ell_{\textrm{hybrid}}$, at which
  $\beta_{\ell}=0.15$. Bottom panel: The ratio between the transfer
  functions in the top panel. Note the excellent agreement up to
  $\ell\approx 800$, after which the differences in pixel window
  approximations becomes visible.}
\label{fig:gauss_comp}
\end{figure}

\begin{figure}[t]
\mbox{\epsfig{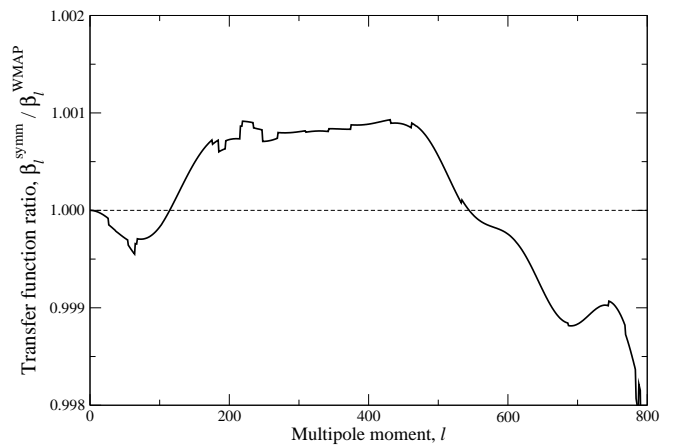}}
\caption{Comparison of the beam transfer functions derived from our
  symmetrized beam profile of the V1 differencing assembly with the
  official WMAP beam transfer function. The differences between the
  two symmetrized transfer functions are generally smaller than 0.1\%,
  indicating that the inputs to our analysis are consistent with the
  WMAP inputs.}
\label{fig:symmetrized_comparison}
\end{figure}

\begin{deluxetable*}{lccccccc}
\tablecaption{Summary of DA parameters\label{tab:summary}}
\tablecolumns{8}
\tablehead{DA  & FWHM\tablenotemark{a} & Radius\tablenotemark{b} &
  $N_{\textrm{side}}$ & $\ell_{\textrm{max}}$ &
  $\ell_{\textrm{hybrid}}$ & $N_{\textrm{samples}}$ &$\sigma_{\textrm{N}}$\tablenotemark{c}\\
& (arcmin) & (degrees) & & & & $(10^8)$& $(\mu\textrm{K})$}
\startdata
K1   & 53 & $7.0$ & 512  & 750  & 318       & 2.5 & \nodata \\
Ka1  & 40 & $5.5$ & 512  & 850  & 411       & 2.5 & \nodata \\
Q1   & 31 & $5.0$ & 512  & 1100 & 522       & 3.1 & 78.2\\
Q2   & 31 & $5.0$ & 512  & 1100 & 515       & 3.1 & 74.2 \\
V1   & 21 & $4.0$ & 1024 & 1500 & 789       & 4.1 & 99.0\\
V2   & 21 & $4.0$ & 1024 & 1500 & 779       & 4.1 & 88.2\\ 
W1   & 13 & $3.5$ & 1024 & 1700 & 1164      & 6.2 & 143.8\\
W2   & 13 & $3.5$ & 1024 & 1700 & 1148      & 6.2 & 159.7\\
W3   & 13 & $3.5$ & 1024 & 1700 & 1162      & 6.2 & 168.5\\
W4   & 13 & $3.5$ & 1024 & 1700 & 1169      & 6.2 & $\,\,\,\,\,\,$164.4
\enddata
\tablenotetext{a}{Effective symmetrized beam size.}
\tablenotetext{b}{Radius used for pixelized beam convolutions. See
  \citet{hill:2009} for details. }
\tablenotetext{c}{Average full-sky RMS values evaluated at
  $N_{\textrm{side}}=512$. }
\end{deluxetable*}

We begin by defining our notation. We will be estimating the product
of the WMAP beam transfer function, $b_{\ell}$, and pixel window,
$p_{\ell}$, by direct simulation. This product is denoted
$\beta_{\ell} = b_{\ell} p_{\ell}$. Given this function, the combined
effect on a sky map, $T(\hat{n})$, of convolution by an instrumental
beam and averaging over finite-sized pixels may be approximated in
harmonic space as
\begin{equation}
T(\hat{n}) = \sum_{\ell=0}^{\ell_{\textrm{max}}} \sum_{m =
  -\ell}^{\ell} \beta_{\ell} a_{\ell m} Y_{\ell m}(\hat{n}),
\label{eq:analytic_convolution}
\end{equation}
where $Y_{\ell m}(\hat{n})$ are the usual spherical harmonics.

The angular power spectrum of $T$ is given by
\begin{equation}
\hat{C}_{\ell} = \frac{1}{2\ell+1}\sum_{m=-\ell}^{\ell} \beta_{\ell}^2
|a_{\ell m}|^2,
\end{equation}
while the power spectrum of the true underlying CMB map, $s(\hat{n})$, is
\begin{equation}
C_{\ell} = \frac{1}{2\ell+1}\sum_{m=-\ell}^{\ell} |a_{\ell m}|^2.
\end{equation}
The effect of the beam convolution and pixel averaging on the power
spectrum is therefore simply given by a multiplication with
$\beta_{\ell}^2$.

The overall approach for estimating $\beta_{\ell}$ used in this paper
may be summarized by the following steps: First, we simulate
time-ordered data (TOD) for each DA, taking into account both the
detailed beam maps of WMAP and the exact orientation of the spacecraft
at each point in time. We then produce a sky map from this TOD. Next
we compute the square root of the ratio between the output and the
input power spectra. Finally, because the input beam maps themselves
are pixelized, and therefore slightly smoothed, we also have to
divide out (or deconvolve) the pixel window of the beam
pixelization. The resulting function becomes our estimate of the beam
transfer function, $\beta_{\ell}$.

\begin{figure*}[t]
\mbox{\epsfig{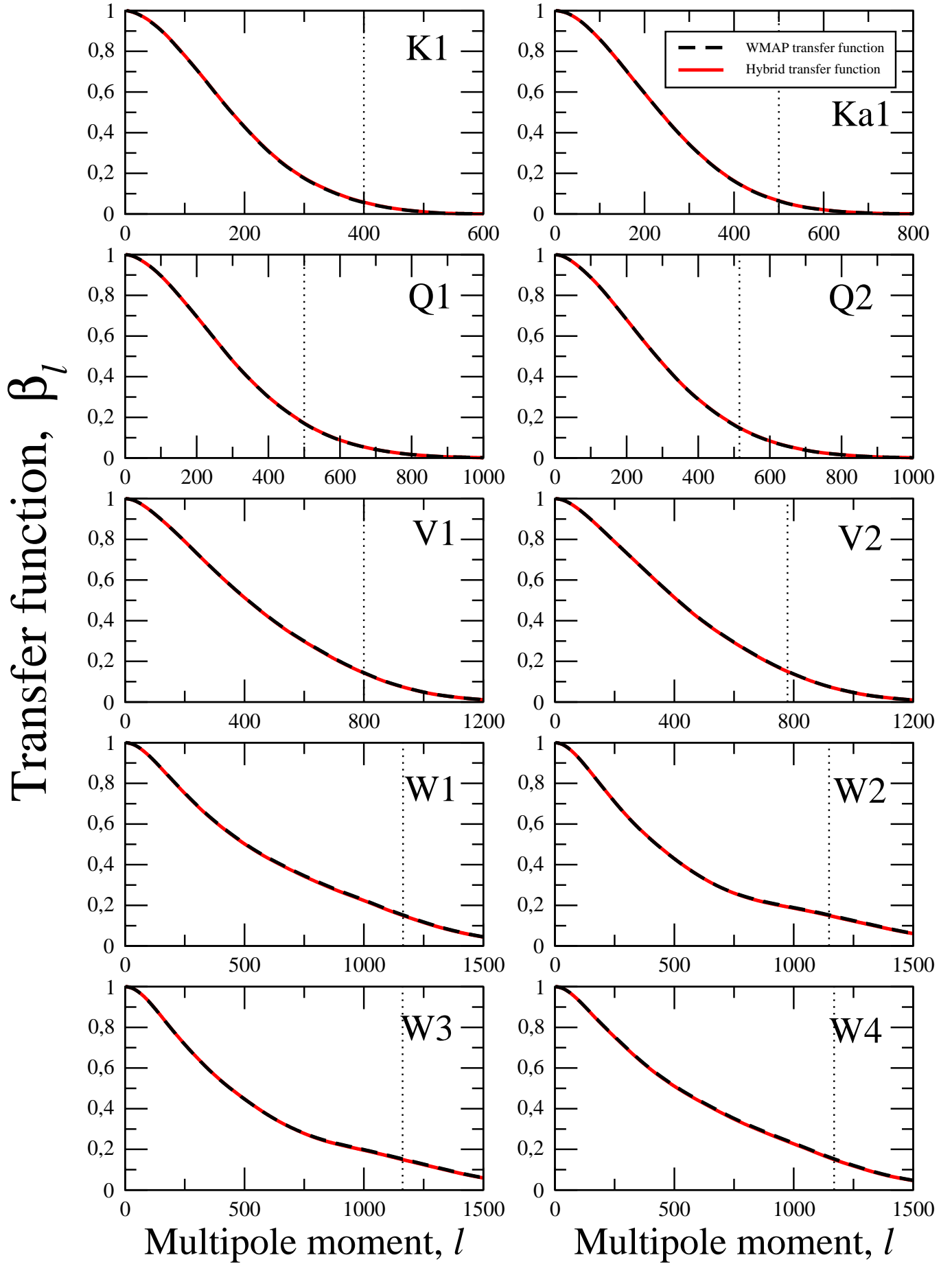}}
\caption{Comparison between transfer functions derived in this paper
  (red lines) to the nominal WMAP transfer functions (black dashed
  lines). The transition multipole, $\ell_{\textrm{hybrid}}$ is marked
by dotted vertical lines.}
\label{fig:beam_comp}
\end{figure*}

\begin{figure*}[t]
\mbox{\epsfig{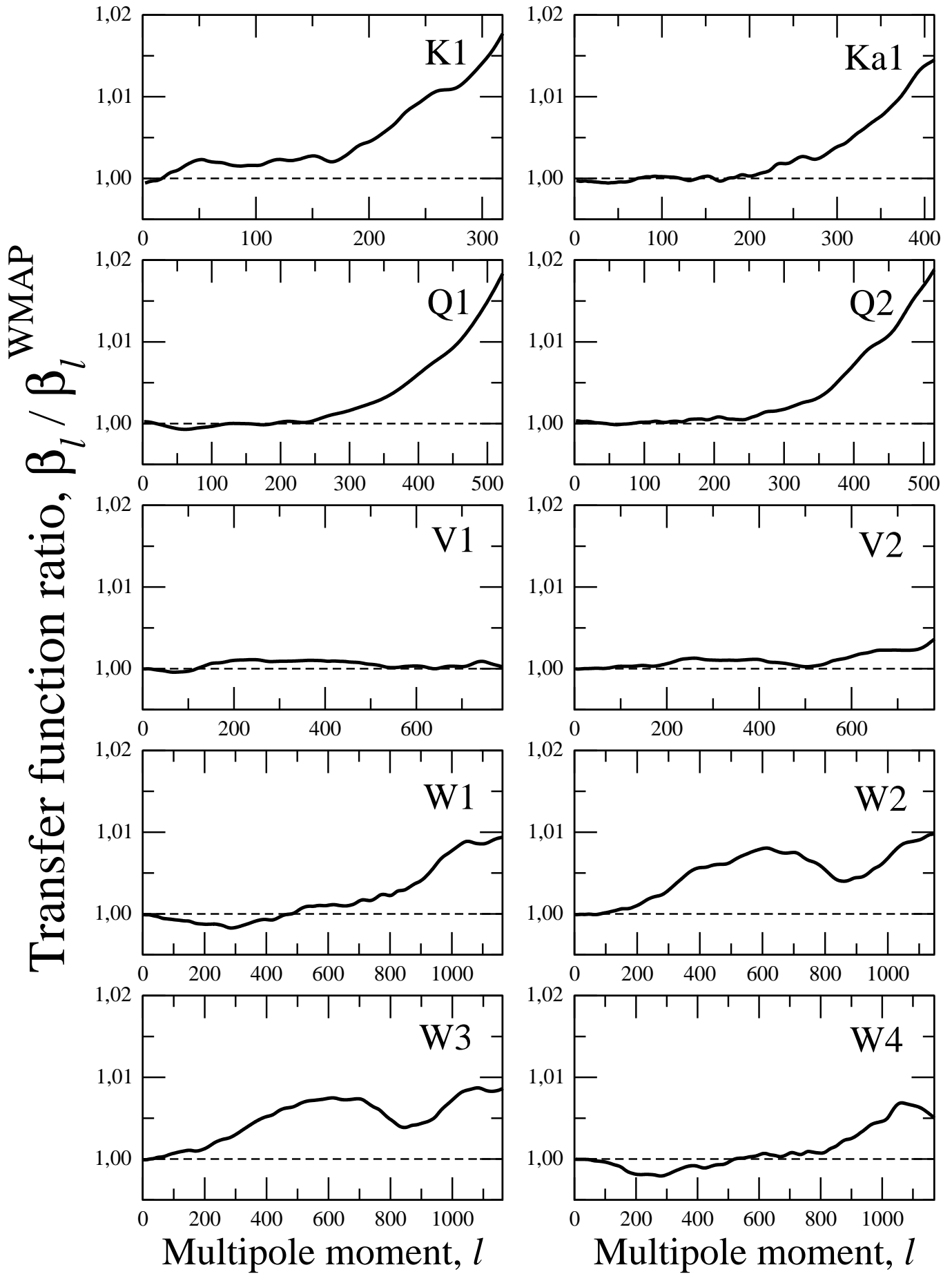}}
\caption{The ratio between the transfer functions derived in this
  paper and the nominal WMAP transfer functions for all DAs. Note that
  the DAs split into two main groups depending on focal plane
  position: The outer DAs, K1, Ka1 and Q1--2, all rise with $\ell$,
  whereas the inner DAs, V1--2 and W1--4, decrease with $\ell$. Note
  also the similarity between W1 and W4, between W2 and W3, and
  between V1 and V2.}
\label{fig:beam_ratios}
\end{figure*}

Note that in this paper we are only concerned with the effect of
asymmetric beams, not other systematic effects such as instrumental
noise. All following discussions will therefore assume noiseless
observations.

\subsection{Simulation of time-ordered data}
\label{sec:simulation}

Our first step is to simulate a reference CMB sky realization, $s$,
given an angular temperature power spectrum,
$C_{\ell}^{\textrm{theory}}$. This can be achieved with a standard
code such as ``anafast'', which is available in the
HEALPix\footnote{http://healpix.jpl.nasa.gov} software package. Note
that this map should not be smoothed with either an instrumental beam
or a pixel window; adding these effects is the task of the following
pipeline. Explicitly, the input reference map should simply be pure
spherical harmonic modes projected onto a set of pixel centers.

Next, we need to be able to convolve this map with a given beam map at
arbitrary positions and orientations on the sphere. In this paper we
do this by brute-force integration in pixel space. For an alternative
fast Fourier space based approach to the same problem, see
\citet{wandelt:2001}.

We define $\hat{p}$ to be a unit vector pointing towards the beam
center, and specify its position on the sphere using longitude and
co-latitude $(\phi,\theta)$. We further define $\psi$ to be the angle
between some fixed reference direction in the beam map and the local
meridian. The value of the beam map at position $\hat{n} =
(\phi',\theta')$, which in principle is non-zero over the full sky, is
denoted $b(\phi', \theta'; \phi, \theta, \psi)$. With these
definitions, the desired convolution may be written as
\begin{equation}
T(\phi, \theta, \psi) = \int_{4\pi} s(\phi',\theta') b(\phi',\theta';
\phi, \theta, \psi) d\Omega'.
\label{eq:convolution}
\end{equation}

Computationally speaking, we approximate this integral as a direct sum
over HEALPix pixels, which all have equal area, with the product
$s\cdot b$ being evaluated at HEALPix pixel centers. To make these
calculations computationally feasible, we assume that the beam is zero
beyond some distance from the beam center (ranging between 3.5 and
$7^{\circ}$ for the WMAP channels), and thus only include the main
lobe in the following analysis. While the WMAP beam maps are provided
as pixelized maps, we need to know the beam values at arbitrary
positions (ie., HEALPix pixel centers). We solve this by computing a
2D spline for each beam map, enabling us to interpolate to arbitrary
positions. For a review on one specific method for fast (and local) 2D
spline evaluations, see Appendix \ref{sec:splines}.

WMAP is a differential experiment, and measures at each point in time
the difference between the signals received by two different
detectors, denoted A and B. The full set of time-ordered WMAP data may
therefore be written as
\begin{equation}
d_x(i) = T_x^{\textrm{A}}(i) - T_x^{\textrm{B}}(i),
\end{equation}
where $x = \{\textrm{K1, Ka1, Q1-2, V1-2, W1-4}\}$ is a DA label, and
$i$ is a time index, and for each detector a short-hand for
$(\phi,\theta,\psi)$. This equation may be written in the
following matrix form,
\begin{equation}
\mathbf{d}_x = \mathbf{A} \mathbf{T}_x
\end{equation}
where we have introduced an $N_{\textrm{tod}} \times
N_{\textrm{pix}}$ pointing matrix $\mathbf{A}$. This matrix contains two
numbers per row; 1 in the column hit by the center of beam A at time
$i$, and -1 in the column hit by the center of beam B.

The remaining problem is to determine the position and orientation of
each detector at each time step. This information has been made
publicly available by the WMAP team on
LAMBDA\footnote{http://lambda.gsfc.nasa.gov}, and consists of a large
set of pointing files together with useful IDL routines for extracting
the desired information.

\subsection{Map making with differential data}
\label{sec:map_making}

For the map making step we adopt the algorithm developed by
\citet{wright:1996}, which was used in the 1-year WMAP pipelines
\citep{hinshaw:2003, hinshaw:2007}. Here we only summarize the
essential algebra, and outline the algorithm.

Our goal is to establish an unbiased and, preferably, optimal
estimate of the (smoothed) sky signal, $\hat{\mathbf{T}}$, given a set
of differential TOD values, $\mathbf{d}$. For noiseless data, the
maximum likelihood estimator is simply 
\begin{equation}
\hat{\mathbf{T}} = (\mathbf{A}^t \mathbf{A})^{-1} \mathbf{A}^t \mathbf{d}.
\label{eq:map_making}
\end{equation}

For high-resolution sky maps, this equation involves an inverse of a
large matrix and cannot be solved explicitly. Instead, one often
resorts to iterative methods such as Conjugate Gradients, or, for
differential data, the method developed by \citet{wright:1996}.

We present the iterative differential map maker in a simple manner:
Define $\mathbf{D}$ to be the diagonal matrix that counts the number
of hits $N_{\textrm{obs}}(p)$ per pixel $p$ on the diagonal, and $a_i$
and $b_i$ to be the pixels hit by side A and side B at time $i$,
respectively. Suppose that we already have established some estimate
for the solution, $\hat{\mathbf{T}}^j$. (Note that this can be zero.) Then
the iterative scheme
\begin{equation}
  \hat{\mathbf{T}}^{j+1} = \hat{\mathbf{T}}^{j} +
  \mathbf{D}^{-1}\mathbf{A}^{t}(\mathbf{d}-\mathbf{A}\hat{\mathbf{T}}^{j})
  \label{eq:iterative_map_maker}
\end{equation}
will converge to the true solution: If $\hat{\mathbf{T}}^{j} =
\mathbf{T}$, then $\mathbf{d} = \mathbf{A}\hat{\mathbf{T}}^{j}$, and the
second term on the right hand side is zero. This algorithm is 
implemented by the following scheme:
\begin{equation}
  \hat{T}^{j+1}_p = \frac{\sum_{i}
  \left(\delta_{p,a_i} \left[\hat{T}^{j}_{b_i} + d_i\right] +
  \delta_{p,b_i}\left[\hat{T}^j_{a_i} - d_i \right] \right)}{N_{\textrm{obs}}(p)}.
\end{equation}

This algorithm was originally presented by \citet{wright:1996}. The only
new feature introduced here is the choice of starting point. In the
original paper, \citet{wright:1996} initialized the iterations at the DMR
dipole, since their test simulation included a CMB dipole
term. However, for a given scanning strategy, there will often be some
large-scale modes that are less well sampled than others. For
instance, for the WMAP strategy $\ell=5$ is more problematic than
other modes \citep{hinshaw:2003}. This leads to slow convergence with
the above scheme for this mode. 

We therefore choose a different approach: Before solving for the
high-resolution map by iterations, we solve Equation
\ref{eq:map_making} by brute-force at low resolution. For the cases
considered later in this paper, we choose a HEALPix resolution of
$N_{\textrm{side}}=16$ for this purpose. With 3072 pixels, about 30
seconds are needed to solve this system by singular value
decomposition. (Note that the monopole is arbitrary for differential
measurements, and one must therefore use an eigenvalue decomposition
type algorithm to solve the system.) The improvement in convergence
speed due to this choice of initial guess is explicitly demonstrated
in Appendix \ref{sec:convergence}.

Our convergence criterion is chosen such that the RMS difference
between two consecutive iterations must be less than 0.05 $\mu$K. We
have verified that this leads to errors of less than 0.1 $\mu$K in the
final solution, of which far most is due to a residual dipole. This is
typically achieved with 30--50 iterations, although some converge
already after 20--30 iterations and a few after 70 or more iterations.

At first glance, the fact that the final residuals are as small as 0.1
$\mu$K for an RMS stopping criterion as large as 0.05 $\mu$K may seem
surprising. However, this is explained by the fact that the iterative
solution obtained by Equation \ref{eq:iterative_map_maker} often
alternates between high and low values about the true answer. This
suggests that a further improvement to the algorithm may be possible:
Faster convergence may perhaps be obtained by computing the average of
two consecutive iterations, $\hat{\mathbf{T}} = (\hat{\mathbf{T}}^{j}
+ \hat{\mathbf{T}}^{j+1})/2$, as the map estimate for iteration
$j+2$. However, the computational resources spent during map making is
by far sub-dominant compared to the TOD simulation, and we have
therefore not yet implemented this step in our codes.

\subsection{Estimation of hybrid beam transfer functions}
\label{sec:transfer_functions}

As described in the introduction to this section, we estimate the
transfer function by the square root of the ratio between the power
spectra of the convolved map and the input map
\begin{equation}
\hat{\beta}^2_{\ell} = \sqrt{\frac{\hat{C}_{\ell}}{C_{\ell}}}.
\end{equation}
However, as noted above, this function describes both the effect from
instrumental beam smoothing and averaging over pixels. In the present
paper we are concerned mostly with the former of these, which has a
stronger impact on large to intermediate scales. 

In the following we choose to construct a hybrid transfer function,
\begin{equation}
\hat{\beta}_{\ell} = \left\{
\begin{array}{lcl}
\sqrt{\frac{\hat{C}_{\ell}}{C_{\ell}}} & \textrm{for} &\ell \le\ell_{\textrm{hybrid}} \\
\sqrt{\frac{\hat{C}_{\ell_{\textrm{hybrid}}}}{C_{\ell_{\textrm{hybrid}}}}}
\frac{b_{\ell}^{\textrm{WMAP}}}{b_{\ell_{\textrm{hybrid}}}^{\textrm{WMAP}}}
\frac{p_{\ell}}{p_{\ell_{\textrm{hybrid}}}}& \textrm{for} & \ell > \ell_{\textrm{hybrid}} 
\end{array}
\right..
\end{equation}
Here $b_{\ell}^{\textrm{WMAP}}$ is the nominal symmetrized transfer
function published by the WMAP team, $p_{\ell}$ is the (uniformly
averaged) HEALPix pixel window, and $\ell_{\textrm{hybrid}}$ is some
transition multipole. In other words, we adopt our own direct estimate
of the transfer function up to $\ell_{\textrm{hybrid}}$, but the
symmetrized, asymptotically uniform and properly scaled WMAP transfer
function at higher multipoles.

Note that this issue is of minor importance in terms of cosmological
interpretation, i.e., angular power spectrum and cosmological
parameters, because the transition typically takes place in the noise
dominated high-$\ell$ regime. The effect of the anisotropic pixel
window is therefore largely suppressed. In the present paper, we
therefore choose to focus on the beam dominated region, and leave a
detailed study of the pixel window to a future paper. 

Finally, because we only generate a relatively small number of
simulations in this paper, there is considerable Monte Carlo scatter
in our estimated transfer functions on an $\ell$-by-$\ell$ basis. To
reduce this Monte Carlo noise, we smooth all transfer functions using
the smooth spline formalism described by, e.g., \citet{green:1994}.

\section{Data and simulations}
\label{sec:data}

All data products used in this study are provided by the WMAP team on
LAMBDA as part of their 5-year data release. However, the calculations
performed here are computationally extremely demanding, and we
therefore include only roughly one year worth of data in our
calculations. To be precise, we include the period between July 10th
2001 and August 2nd 2002, except for three days with missing data, for
a total of 383 days\footnote{Our original intention was to include
  precisely one year of observations in our analysis, and therefore we
  processed 365 WMAP pointing files. However, we noticed after the
  calculations were completed that some of the pointing files
  contained slightly more than one day's worth of data, such that a
  total of 383 days was in fact included.}.

We consider all 10 WMAP DAs in our calculations, which are denoted, in
order from low to high frequencies, K1 (23 GHz), Ka1 (33 GHz), Q1--2
(41 GHz), V1--2 (61 GHz) and W1--4 (94 GHz), respectively. Their
resolutions range from 53' FWHM at K-band to 13' FWHM at
W-band. Because of this large range in resolution, we specify the
pixel resolution and harmonic space range for each case
separately. For instance, K-band is pixelized at
$N_{\textrm{side}}=512$, and includes multipoles up to
$\ell_{\textrm{max}}=750$ (the highest multipole present in the
transfer function provided by the WMAP team), while the W-band is
pixelized at $N_{\textrm{side}}=1024$, and includes multipoles up to
$\ell_{\textrm{max}}=1700$. A full summary of all relevant parameters
for each DA is given in Table \ref{tab:summary}. 

Note that the listed noise RMS values are only used for estimating the
power spectrum weights in Section \ref{sec:cosmology}. For simplicity
we have adopted the official RMS values for the foreground-reduced
5-year WMAP maps here, but note that there is a $\sim1$\% bias in some
of these values \citep{groeneboom:2009b}. However, this has no
significant impact on the results presented in this paper.

\begin{deluxetable}{lccc}
\tablecaption{Cosmological parameters\label{tab:parameters}}
\tablecomments{Comparison of cosmological parameters derived from the
  nominal WMAP power spectrum (second column) and from the power
  spectrum corrected for asymmetric beams (third column). The
  rightmost column shows the relative shift between the two in units
  of $\sigma$.}
\tablecolumns{4}
\tablehead{Parameter          & Nominal WMAP & Corrected beams & Shift in $\sigma$}
\startdata
$\Omega_{\textrm{b}}h^2$        & $0.0228 \pm 0.0006$  & $0.0228 \pm 0.0006$ & $0.0$ \\
$\Omega_{\textrm{cdm}}h^2$      & $0.109 \pm 0.0006$  & $0.109 \pm 0.006$ & $0.0$ \\
$\log(10^{10} A_{\textrm{s}})$  & $3.064 \pm 0.042$  & $3.058 \pm 0.042 $ & $-0.2$  \\ 
$\tau$                        & $0.089 \pm 0.017$  & $0.089 \pm 0.017$ & $0.0$ \\
$h$                           & $0.722 \pm 0.027$  & $0.725 \pm 0.026$ & $0.1$ \\
$n_{\textrm{s}}$                & $0.965 \pm 0.014$  & $0.964 \pm 0.014$ & $-0.1$ 
\enddata
\end{deluxetable}

The beam maps for each DA are provided in the form of pixelized maps,
and separately for side A and B. Each beam map contains non-zero
values inside a radius around the beam center which is specified for
each DA. For instance, the K-band radius is $7^{\circ}$, and the
W-band radius is $3\fdg5$. When evaluating the convolution defined in
Equation \ref{eq:convolution}, we include all pixels inside this
radius.

The pixel size of the beam maps is $2.4'$, which over-samples even the
W-band beams. Based on these high-resolution maps we precompute all
coefficients of the corresponding bi-cubic spline (see Appendix
\ref{sec:splines}), which allows us to very quickly interpolate at
arbitrary positions in the beam map with high accuracy.

The pixel window of the $2.4'$ beam pixelization is also provided on
LAMBDA, and is taken into account by deconvolving final results
whereever appropriate. Note that this effect must be taken properly
into account also by other users who wish to use our simulations for
follow-up studies.

Each beam is normalized by convolving a map constant equal to 1 at
1000 random random positions and orientations, and demanding that the
average of the resulting 1000 values equals unity. With the 2D spline
interpolation scheme described in Appendix \ref{sec:splines} the
random uncertainties on the normalization due to beam position and
orientation are $\sim0.2$\%. For comparison, directly reading off
pixel values from the beam maps without interpolation leads to
variations in the normalization at the $\sim2$\% level.

For our base CMB reference sky set, we draw ten random Gaussian
realizations from the the best-fit $\Lambda$CDM power spectrum derived
from the 5-year WMAP data alone \citep{komatsu:2009}. These maps are
generated at both $(N_{\textrm{side}}=512$ and
$N_{\textrm{side}}=1024$ using the same seeds, and include neither an
instrumental beam nor a pixel window; they are simply spherical
harmonic modes projected onto the HEALPix pixel centers. All ten
realizations are processed for all ten DAs, such that the resulting
simulations may be used for multifrequency analysis, if so desired.

\begin{figure*}[t]
\begin{center}
\mbox{\epsfig{figure=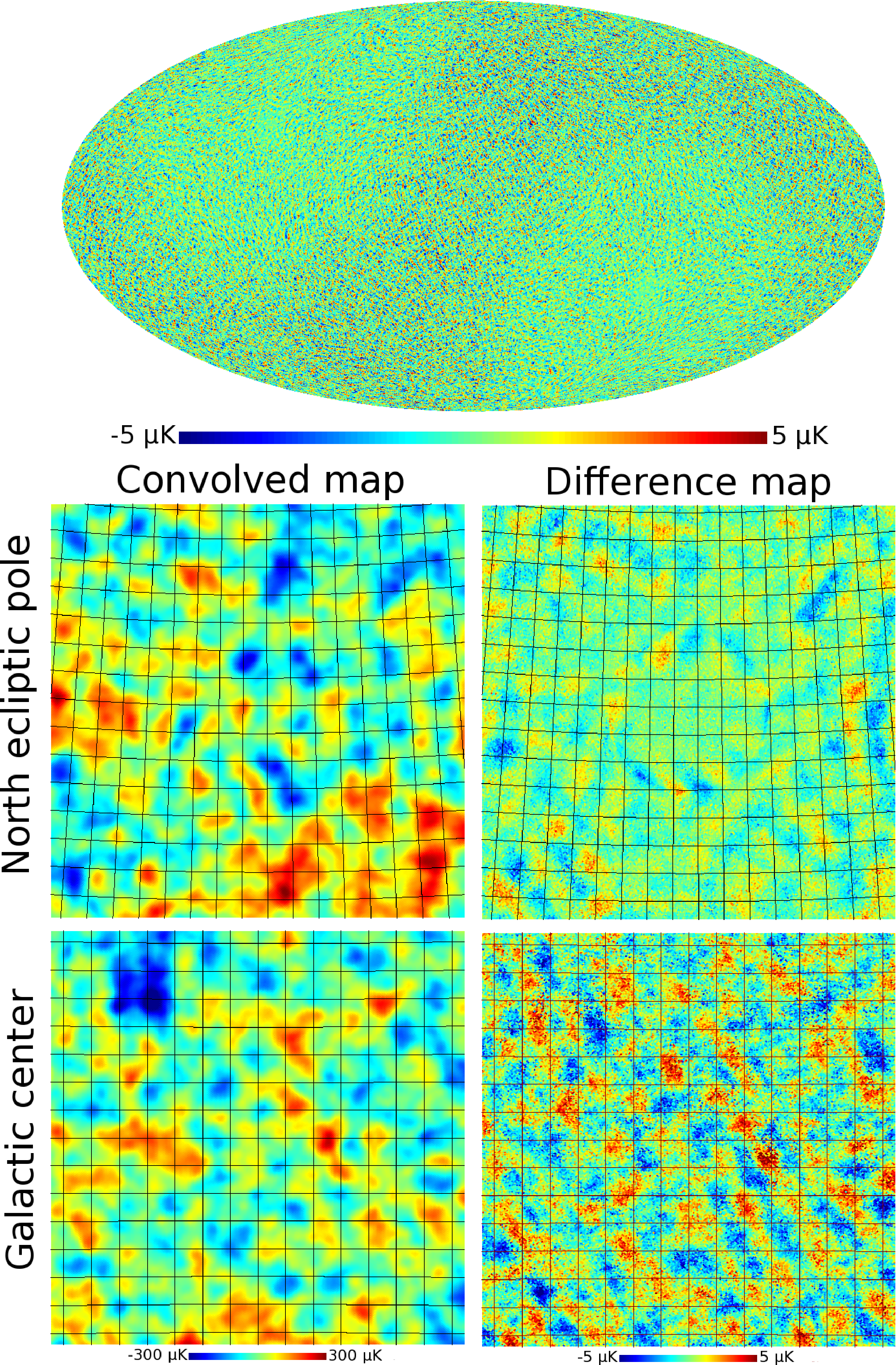,width=0.8\linewidth,clip=}}
\end{center}
\caption{Top panel: Difference between a V1 simulation convolved with
  the full asymmetric beam and the same realization convolved with
  the corresponding symmetrized transfer function. The two maps have
  identical power spectrum but different phases. Note that larger
  differences are observed along the ecliptic plane, where the density
  of observations is lower than towards the ecliptic poles, and the
  cross-linking is also weaker. Bottom panels: Zoom-in on two regions,
the north ecliptic pole (NEP; top row) and the Galactic center (GC;
bottom row). Left column shows the map convolved with an asymmetric
beam, and right column shows the same difference as in the top panel.}
\label{fig:asym_vs_sym}
\end{figure*}

As noted above, the computational requirements for the analyses
presented here are demanding. The CPU time for processing a single
W-band DA is $\sim4000$ hours, and the total disk usage for the entire
project is $\sim$1TB. For comparison, the corresponding map making
step requires $\sim60$ CPU hours, and is thus completely sub-dominant
the TOD simulation.

\section{Comparison with analytic case}
\label{sec:analytic_comparison}

In order to test our pipeline and understand its outputs, we start by
considering a perfect Gaussian beam. This case is treated in two
different ways: First, we convolve a CMB realization directly in
harmonic space (as defined by Equation \ref{eq:analytic_convolution})
with a $\sigma_{\textrm{fwhm}}=20'$ FWHM analytic Gaussian beam and
the appropriate HEALPix pixel window, $p_{\ell}$ for
$N_{\textrm{side}}=1024$. The combined transfer function for this case
reads
\begin{equation}
\beta_{\ell}^{\textrm{ref}} = e^{-\frac{1}{2}\ell(\ell+1) \sigma^2}
p_{\ell},
\label{eq:gaussian_beam}
\end{equation}
where $\sigma = \sigma_{\textrm{fwhm}} / \sqrt{8\ln 2}$, and
$\sigma_{\textrm{fwhm}}$ is expressed in radians.

Second, we map out a corresponding two-dimensional Gaussian in pixel
space over a grid of $2.4'$ pixels, the same resolution as the WMAP
beam maps. We then input this into our simulation pipeline together
with the same CMB realization used for the analytic convolution, and
with the V1 channel pointing sequence. From the resulting brute-force
convolved map, we then obtain the effective transfer function,
$\beta_{\ell}$, as described in Section \ref{sec:transfer_functions}.

This function is plotted in the top panel of Figure
\ref{fig:gauss_comp}, together with the product of the analytic
Gaussian beam and the HEALPix pixel window. The ratio of the two
effective functions is shown in the lower panel. 

From this figure it is clear that the agreement between the two
approaches is excellent up to $\ell \approx 800$. At higher $\ell$'s,
however, the ratio increases rapidly, indicating that the analytic
approach smooths more than the brute-force approach. 

In practice, one is well advised not to consider scales smaller than
those that are properly oversampled by the scanning strategy. In this
paper, we adopt the analytic case considered in this section to guide
us in determining which scale that is. Explicitly, we conservatively
demand that that the effective beam transfer function must be greater
than 0.15 in order to consider it to be properly oversampled, and
therefore independent of scanning strategy. We adopt the corresponding
multipole moment to be $\ell_{\textrm{hybrid}}$, as defined in Section
\ref{sec:transfer_functions}. Thus, the symmetrized WMAP beam and
HEALPix pixel window are used at scales for which the beam amplitude
drops below 0.15.

This test shows that the computational machinery described in Section
2 works as expected. However, it does not validate the
inputs. Therefore, to check that our input data are consistent with
those used in the official WMAP analysis
\citep{page:2003,jarosik:2007,hill:2009}, we repeat their approach of
Legendre transforming a symmetrized beam profile,
$b_{\textrm{s}}(\theta)$ into the harmonic space transfer function,
$b_{\ell}$, but now derive $b_{\textrm{s}}(\theta)$ from our 2D beam
splines. 

The result from one of these computations is shown in Figure
\ref{fig:symmetrized_comparison}. Here we see that the agreement is
better than 0.1\% over most of the range.

\section{The effect of asymmetric beams in WMAP}
\label{sec:WMAP}

We now present the main results obtained in this paper, namely the
effective beam transfer functions for each WMAP DA, taking into
account both the full asymmetric beam patterns and scanning
strategy. These are shown in Figure \ref{fig:beam_comp} (red lines),
and compared to the nominal WMAP transfer functions (dashed black
lines). The vertical dotted line indicates $\ell_{\textrm{hybrid}}$
for each case.

Clearly, the differences between the two sets of results are small, as
no visual discrepancies are seen in this plot. In Figure
\ref{fig:beam_ratios} we plot the ratio between our transfer functions
and the WMAP transfer functions for $\ell \le \ell_{\textrm{hybrid}}$,
and now we do see small but significant differences between the two
sets of results.

Before looking at the results, we note that the completely symmetrized
transfer function constitutes a lower bound on the full transfer
function: When symmetrizing the beam, any beam mode with $m \ne 0$ is
nullified. Consequently, less power is retained after convolution with
the symmetrized beam than when an arbitrary beam is considered.

In Figure \ref{fig:beam_ratios} we see precisely this
behaviour. Explicitly, we see that the ratios are essentially unity on
the largest scales (smallest $\ell$'s), and then increase to higher
$\ell$'s. (Some functions show values slightly lower than unity over
short ranges, typically $\lesssim$0.2\%, and this is due to small
differences in two beam models used by WMAP and our analysis; similar
differences are observed when comparing symmetrized transfer
functions.) 

The point at which the two transfer functions start to diverge varies
somewhat from DA to DA, and depends of course on the angular scale of
the particular DA. For instance, W2 and W3 start to diverge already
$\ell\sim200$, while W1 and W4 are very close up to $\ell \sim
800$. On the other hand, all the low-frequency DAs are generally close
to unity up to $\ell \sim 200-300$,

These general and qualitative remarks reflects the position of each DA
in the WMAP focal plane (see Figure 6 of Jarosik et al.\ 2007 for an
excellent visualization of the A side beams): K1, Ka1 and Q1--2 are
positioned the furthest away from the optical axis, while V1--2 and
W1--4 are the closest. Similarly, W1 and W4 are positioned lower in
elevation, and generally have more sub-structure, than W2 and W3.

However, it should be emphasized that the overall differences are
small, typically less than than 1\% at
$\ell\le\ell_{\textrm{hybrid}}$. Further, these differences are only
significant (again, with the exception of W1 and W4) in the
intermediate- and high-$\ell$ ranges.

\begin{figure}[t]

\mbox{\epsfig{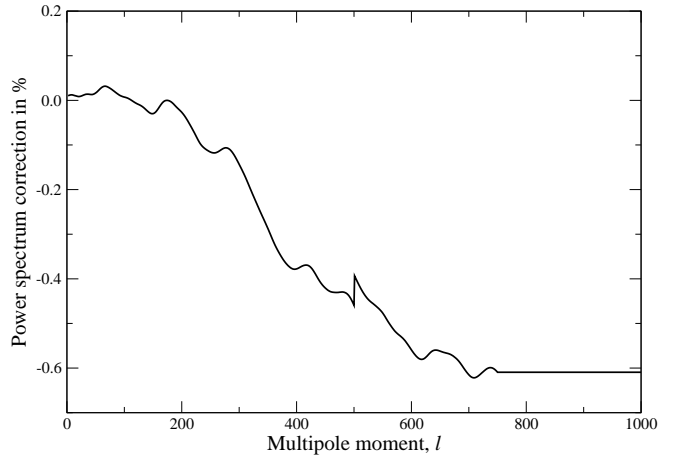}}

\caption{Total correction to the 5-year co-added WMAP temperature
  power spectrum due to asymmetric beams. Note the transition between
  high and low signal-to-noise weighting schemes at $\ell=500$, and
  also the manually capped amplitude at $\ell>750$. The latter is
  imposed in order to be conservative in the very high-$\ell$ regime,
  where the transfer functions are sensitive to pixel window effects.}
\label{fig:power_corr}
\end{figure}

To build up some intuitive understanding of the spatial variations
caused by the asymmetric WMAP beams, we show in Figure
\ref{fig:asym_vs_sym} the difference between the fully asymmetrically
convolved map and the corresponding map convolved with the symmetrized
transfer function directly in harmonic space for one of the V1
simulations. Thus, the two convolved maps have identical power
spectra, but slightly different phases. The top panel shows the
full-sky difference map with a temperature scale of
$\pm5\mu\textrm{K}$. The lower panels show two selected
$15^{\circ}\times15^{\circ}$ regions centered on the north ecliptic
pole (NEP; top row) and the Galactic center (GC; bottom row),
respectively. The left column shows the actual temperature map
convolved with the asymmetric beam, and the right column shows the
same differences as in the full-sky plot.

The first striking feature seen in this map is that the differences
are clearly larger in the ecliptic plane than around the ecliptic
poles. This is due to the WMAP scanning strategy, which leads to a
larger number of observations per pixels around the poles, and also
with a greater range of beam orientations. Next, it is difficult to
spot any single unambiguous and well-defined correlation between the
convolved and the difference maps. Clearly, there are similar
morphological structures in the two, but the sign of the correlations
appears to vary. Third, we see a clear tendency of diagonal striping
in the GC plot, which corresponds to correlations along ecliptic
meridians and lines of constant latitude. (Note that these plots are
shown in Galactic coordinates, while the WMAP scanning strategy is
nearly azimuthally symmetric in ecliptic coordinates.)

In the next section, we consider the impact of the asymmetric beams on
cosmological parameters. However, before concluding this section we
make a comment concerning an outstanding issue regarding the 3-year
WMAP power spectra first noted by \citet{eriksen:2007}. They pointed
out the presence of a $3\sigma$ amplitude discrepancy between the V-
and W-band power spectra (Figure 5 of Eriksen et
al.\ 2007). Specifically, the V-band spectrum was biased low compared
to the W-band spectrum between $\ell=300$ and 600 by
$\sim80\mu\textrm{K}^2$. \citet{huffenberger:2006} later showed that
$\sim30\mu\textrm{K}^2$ of this discrepancy could be attributed to
over-estimation of point source power in the 3-year WMAP spectrum
analysis, and this was subsequently confirmed and corrected by
\citet{hinshaw:2007}. Still, about $50\mu\textrm{K}^2$ of this
difference remained, which was statistically significant at $\sim
2\sigma$.

\citet{eriksen:2007} proposed that this difference might be due to
errors in the beam transfer functions caused by asymmetric
beams. Given the new results presented in this paper, we are now in a
position to consider this issue more quantitatively. The relevant
question is then whether the WMAP V-band transfer functions are
systematically biased high compared to the W-band functions. At first
glance, one may get this impression from the plots shown in Figure
\ref{fig:beam_ratios}: The V-band ratios are both very close to unit
over the full range, whereas W2 and and W3 are slightly high in the
same range, at about 0.5\%. On the other hand, W1 and W4 are also very
close to zero in the same range. The net difference is therefore not
more than a few tenths of a percent, which corresponds to
$\sim10\mu\textrm{K}^2$ in the power spectrum. Thus, it is possible
that this effect may contribute somewhat to the power spectrum
discrepancy between V- and W-band, but it does not seem to fully
explain the difference.

\section{Impact on cosmological parameters}
\label{sec:cosmology}

For completeness, we now assess the impact of asymmetric beams in WMAP
on cosmological parameters. We do this by modifying the co-added
5-year WMAP temperature power spectrum \citep{nolta:2009} provided
with the WMAP likelihood code \cite{dunkley:2009, komatsu:2009}
according to the transfer function ratios shown in Figure
\ref{fig:beam_ratios}, and run CosmoMC \citep{lewis:2002} to estimate
the resulting parameters. Only a simple 6-parameter $\Lambda$CDM model
is considered in this paper. For comparison, we also run the code with
the nominal WMAP spectrum as input, so that we can directly estimate
the impact of asymmetric beams with everything else held fixed.

Unfortunately, the individual cross-spectra for each pair of DAs have
not yet been published by the WMAP team, but only the total co-added
spectrum. We must therefore make a few approximations in order to
apply the proper beam corrections to the full spectrum. First, let
$\sigma_n^i$ denote the white noise level of DA $i$ (see Table
\ref{tab:summary}), $\beta_{\ell}^i$ the transfer function estimate
derived in this paper, and let $\beta_{\ell}^{i,\textrm{WMAP}}$ be the
nominal WMAP transfer function. Finally, define
\begin{equation} 
  \delta_{\ell}^i = \frac{\beta_{\ell}^i}{\beta_{\ell}^{i,\textrm{WMAP}}} - 1
\end{equation}
to be the fractional difference between the two.

Next, the WMAP team uses the MASTER pseudo-spectrum algorithm
\citep{hivon:2002} for power spectrum estimation \citep{hinshaw:2003,
  hinshaw:2007, nolta:2009}, which quickly produces good estimates at
high $\ell$'s. However, this method is not a maximum-likelihood
estimator, and it does not yield optimal error bars. To improve on
this, the WMAP applies different pixel weights in different multipole
regions: At low $\ell$'s, where the sky maps are signal dominated,
they apply equal weights to all pixels, while at high $\ell$'s, where
the maps are noise dominated, they apply inverse noise variance pixels
weights. These weights are then taken into account when co-adding the
cross-spectra obtained from all possible DA pairs (but excluding
auto-correlations). The transition is made at $\ell=500$.

The beam-convolved (but noiseless) power spectrum
$\tilde{C}_{\ell}^{ij}$ observed by a given DA pair, $i$ and $j$, may
be written as $\tilde{C}_{\ell}^{ij} = \beta_{\ell}^i \beta_{\ell}^j
C_{\ell}$, where $C_{\ell}$ is the true power spectrum of our sky, and
$\beta_{\ell}^i$ is the true transfer function for DA $i$. The noise
amplitude of the same spectrum is proportional to $\sigma_n^i
\sigma_n^j / \beta_{\ell}^{i}\beta_{\ell}^j$. The inverse noise
variance weight of this cross-spectrum is therefore approximately
\begin{equation}
w_{ij} = \frac{\frac{\beta_{\ell}^i \beta_{\ell}^j}{\sigma_n^i
    \sigma_n^j}}{\sum_{i'<j'}\frac{\beta_{\ell}^{i'} \beta_{\ell}^{j'}}{\sigma_n^{i'}
    \sigma_n^{j'}}}, 
\label{eq:weights}
\end{equation}
 where the sum runs over all $N$ different pairs of
 cross-spectra. (Note that this is only an approximation to the exact
 expression, because other effects also enter the full
 calculations. One important example is the sky cut, which couples
 different $\ell$ modes, and is taken into account through a coupling
 matrix. Such effects are not included in the analysis presented
 here. 

Pulling all of this together, the appropriately co-added power
spectrum provided by WMAP should ideally read
\begin{equation}
\hat{C}_{\ell} = \left\{
\begin{array}{ccc}
\frac{1}{N} \sum_{i < j} \frac{\tilde{C}_{\ell}^{ij}}{\beta_{\ell}^{i}
  \beta_{\ell}^{j}} & \textrm{for} & \ell \le 500 \\
\sum_{i < j} w_{ij} \frac{\tilde{C}_{\ell}^{ij}}{\beta_{\ell}^{i}
  \beta_{\ell}^{j}} & \textrm{for} & \ell > 500
\end{array}
\right..
\label{eq:coadded_spectrum}
\end{equation}
However, the spectrum that in fact is provided by WMAP is Equation
\ref{eq:coadded_spectrum} evaluated for $\beta_{\ell} =
\beta_{\ell}^{\textrm{WMAP}}$, which, according to our calculations,
is slightly biased. To obtain the appropriate correction factor,
$\alpha_{\ell} = \hat{C}_{\ell}/\hat{C}_{\ell}^{\textrm{WMAP}}$, for
each $\ell$, we therefore set $\beta_{\ell} =
\beta_{\ell}^{\textrm{WMAP}} (1+\delta_{\ell})$ in Equation
\ref{eq:coadded_spectrum}, and expand to first order in
$\delta_{\ell}$. Doing this, we find that
\begin{equation}
\alpha_{\ell} = \left\{
\begin{array}{lcc}
1 - \frac{1}{N} \sum_{i < j} (\delta_{\ell}^{i} + \delta_{\ell}^{j}) & \textrm{for} & \ell \le 500 \\
1 - \sum_{i < j} w_{ij}^{\textrm{WMAP}}(\delta_{\ell}^i + \delta_{\ell}^j) & \textrm{for} & \ell > 500
\end{array}
\right.
\end{equation}
where $w_{ij}^{\textrm{WMAP}}$ is the expression given in Equation
\ref{eq:weights} evaluated with $\beta_{\ell}^{\textrm{WMAP}}$.

This function is plotted in Figure \ref{fig:power_corr}. Note,
however, that we have capped the function by hand at $\ell=750$ to be
conservative, considering that our V-band transfer functions do not
have support all the way to the maximum multipole used in the WMAP
likelihood code, $\ell_{\textrm{max}}=1000$.

The results from the corresponding CosmoMC analyses are tabulated in
Table \ref{tab:parameters} in terms of marginal means and standard
deviations. Here we see that there are only very small shifts in the
resulting parameters, indicating good stability with respect to beam
asymmetries in WMAP. For example, there is a positive shift of
$0.2\sigma$ in the amplitude of scalar perturbations,
$A_{\textrm{s}}$, and $-0.1\sigma$ in the spectral index of scalar
perturbations.

\section{Conclusions}
\label{sec:conclusions}

This paper has two main goals. First, we wanted to generate a set of
WMAP-like simulations that fully take into account the asymmetric
beams and anisotropic scanning pattern of the WMAP satellite. Such
simulations are extremely valuable for understanding the impact of
beam asymmetries on various statistical estimators and models. One
example of such, which indeed provided us with the initial motivation
for studying this issue, is the anisotropic universe model presented
by \citet{ackerman:2007}, and later considered in detail with respect
to the WMAP data by \citet{groeneboom:2009a}. The result from that
analysis was a tentative detection of violation of rotational
invariance in the early universe, or some other effect with similar
observational signatures, at the $3.8\sigma$ confidence level. It was
shown that neither foregrounds nor correlated noise could generated
this signal, but the question of asymmetric beams was left
unanswered. This issue will now be revisited in an upcoming paper,
using the simulations generated here.

The second goal of the paper was to assess the impact of beam
asymmetries on the WMAP power spectrum and cosmological parameters. We
did this by comparing the power spectrum of the full beam convolved
simulations with the power spectrum of the input realizations, thereby
providing a direct estimate the effective beam transfer
functions. Doing so, we found differences at the $0.5$\% level in
several differencing assemblies at intermediate and high $\ell$'s with
respect to the nominal WMAP transfer functions. 

A similar analysis was performed for the 3-year WMAP data release by
\citet{hinshaw:2007}, who approach the problem from an analytical
point of view. However, at that time only the A-side beams were
available \citep{hill:2009}, and they therefore assumed identical
beams on both the A and B sides. With this data, they concluded that
the impact of beam asymmetries was $\lesssim 1$\% everywhere below
$\ell = 1000$ for the V- and W-band DAs, in good agreement with our
findings. 

As far as cosmological parameters go, the impact of asymmetric beams
is small. Specifically, we find shifts of $0.2\sigma$ in the amplitude
of scalar perturbations, $A_{\textrm{s}}$, and $-0.1\sigma$ in the
spectral index of scalar perturbations, $n_{\textrm{s}}$.

The simulations described in this paper may be downloaded from IKW's
homepage\footnote{http://www.fys.uio.no/$\sim$ingunnkw/WMAP5\_beams}.

\begin{acknowledgements}
  The authors thank Ned Wright for very useful and stimulating
  discussions, which eventually led to the conception of this project,
  and the referee for very useful comments. We also thank Kris
  G\'{o}rski, Sanjit Gupta, Gary Hinshaw and Graca Rocha for
  interesting discussions, and the WMAP team for providing the
  required data for the project, including the beam pixel window
  function. LA thanks the Institute of Theoretical Astrophysics, Oslo,
  for its generous hospitality during her visit to Norway. IKW, HKE
  and NEG acknowledge financial support from the Research Council of
  Norway. The computations presented in this paper were carried out on
  Titan, a cluster owned and maintained by the University of Oslo and
  NOTUR.  Some of the results in this paper have been derived using
  the HEALPix \citep{gorski:2005} software and analysis package.  We
  acknowledge use of the Legacy Archive for Microwave Background Data
  Analysis (LAMBDA). Support for LAMBDA is provided by the NASA Office
  of Space Science.
\end{acknowledgements}

\appendix

\section{Convergence of the differential map maker}
\label{sec:convergence}

As described in Section \ref{sec:map_making}, we introduce one new
step to the differential map making algorithm presented by
\citet{wright:1996}: We initialize the iterations at the exact
solution of Equation \ref{eq:map_making} evaluated at low resolution,
which in this paper is taken to be $N_{\textrm{side}}=16$, with 3072
pixels. 

To demonstrate the improvement in convergence due to this choice of
initialization, we revisit the analytic case considered in Section
\ref{sec:analytic_comparison}, which compared the results from our
simulation pipeline with an exact analytic case, but taking into
account the actual WMAP scanning strategy. 

\begin{figure*}[t]
\mbox{\epsfig{figure=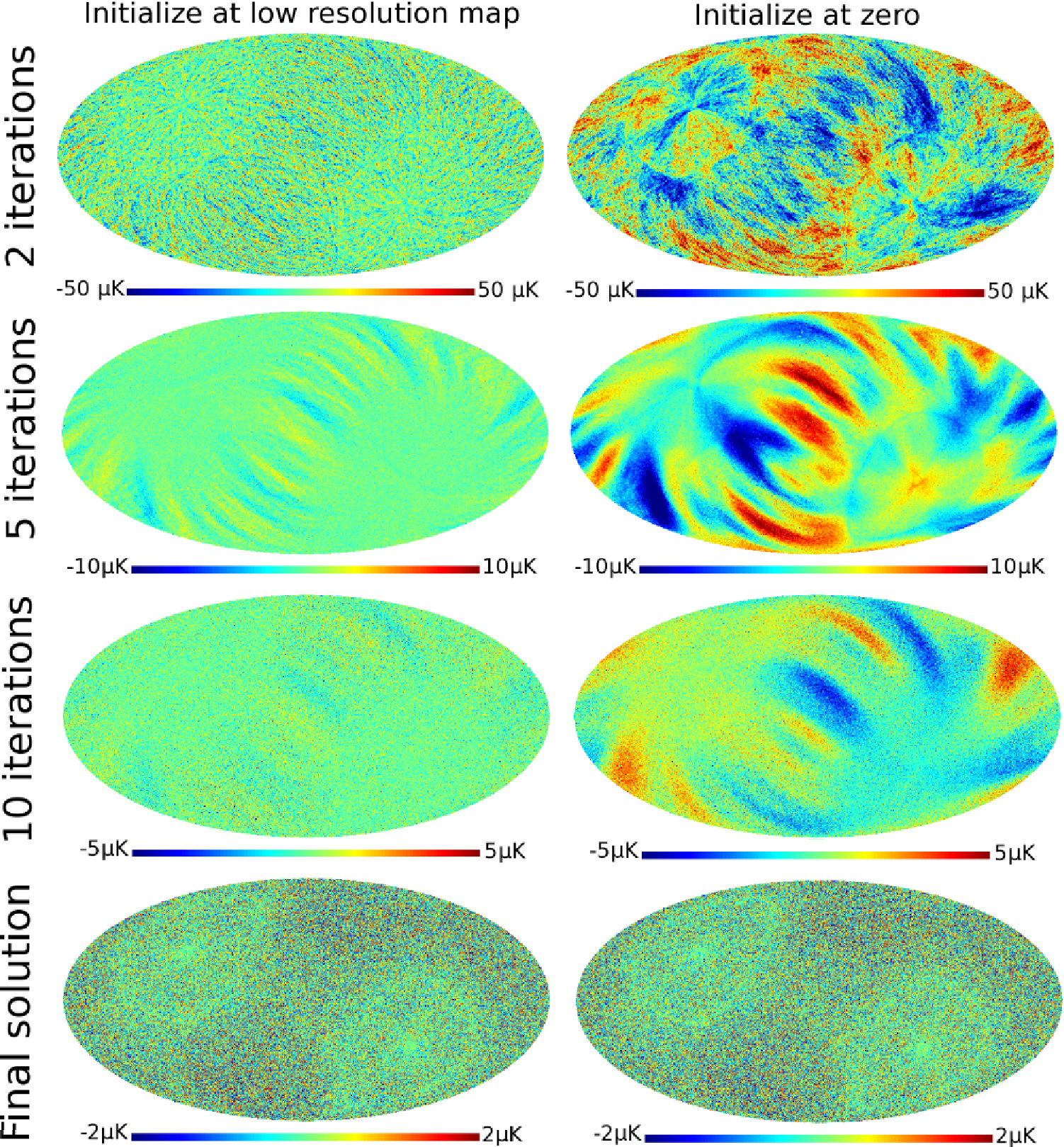,width=\linewidth,clip=}}
\caption{Comparison of convergence of the differential map maker for
  two different choices of initialization. The left column shows the
  snapshots from the series obtained with initializing at a solution
  obtained by brute-force evaluation at low resolution, while the
  right column shows the series obtained when initializating at
  zero. Each plot is a difference map between the current solution for
  a data set including asymmetric beams and real scanning strategy and
  the corresponding map convolved with the analytic Gaussian beam and
  isotropic HEALPix pixel window. The bottom row shows the final
  solutions obtained in the two cases, which were obtained after 67
  and 123 iterations, respectively. These final maps are idential up
  to a $\sim0.1\mu\textrm{K}$ dipole.}
\label{fig:map_iterations}
\end{figure*}

In Figure \ref{fig:map_iterations} we show a set of difference maps
taken between the intermediate solutions produced by the differential
map maker and the analytic and isotropic map solution. From top to
bottom, the panels show the residuals after 2, 5 and 10 iterations,
and at the bottom, the final converged solutions. The left panel shows
the series obtained when initializing the search at the low-resolution
solution, while the right panel shows the series when initializing at
zero. Convergence was achieved respectively after 67 and 123
iterations in the two cases.

Note that the WMAP team initializes their search at the CMB dipole,
which is the dominant component in their data set. However, this is in
our setting equivalent to initializing at zero, since our simulation
does not include a dipole.

Taking the difference between the two final solutions, we have
verified that the peak-to-peak residuals in the two maps are less than
0.1 $\mu$K, of which essentially all is concentrated in a single
dipole component. The solution is thus independent of initialization,
and the only difference lies in computational speed. 

Finally, note that even though the two maps are internally
indistinguishable, they are both quite different from the isotropic
reference map. To be precise, the RMS difference between the derived
maps and the isotropic reference map is 0.91 $\mu$K, with a spatial
pattern similar to the overall WMAP scanning pattern.

The cause of these residuals is the differences in the treatment of
the effective pixel windows: The HEALPix pixel window is computed by
uniformly averaging over the full sky, whereas the simulation pipeline
takes into account the actual pointing directions of the
satellite. Sub-pixel variations in the CMB sky therefore leads to
significant differences in the two estimates on small scales. The
effect of such pixel window variations on the 5-year WMAP power
spectrum will be considered in a future paper.

\section{Fast 2D spline evaluation}
\label{sec:splines}

The heart of the simulation pipeline described in Section
\ref{sec:pipeline} is the real-space convolution algorithm defined by
Equation \ref{eq:convolution}. For this operation to be
computationally feasible we have to be able to evaluate the beam
response function quickly at any position. The real beam maps,
however, are provided to us in the form of two-dimensional pixelized
images with relatively coarse resolution. It is therefore necessary to
establish a fast and accurate interpolation scheme.

We adopt a bicubic spline for this purpose, and review here one
specific implementation of this concept. Note that most of the
following is standard textbook material \citep[e.g.,][]{press:2002},
and is included here only for easy reference.

Suppose we are given some tabulated two-dimensional function $f(x,y)$
over a regular grid, and want to interpolate at arbitrary positions
$(x_0, y_0)$ within this grid. One particularly appealing approach for
doing so are by means of bicubic splines, which are bi-cubic
polynomials in $x$ and $y$,
\begin{equation}
p(x,y) = \sum_{i=0}^3 \sum_{j=0}^3 a_{ij} x^i y^j.
\end{equation}
The coefficients $a_{ij}$ are defined separately for each grid cell,
and our task is to compute these given the tabulated function
$f(x,y)$. Note that once we have these coefficients, any spline
evaluation will be very fast, since it essentially amounts to
performing a vector-matrix-vector multiplication with a $4\times4$
matrix.

Let us first consider a cell defined over the unit square, having
corners $(x,y) = (0,0)$, $(1,0)$, $(0,1)$ and $(1,1)$. (Note that this
assumption does not imply any restriction of the problem, since any
grid cell in a regular grid may be linearly transformed into the unit
square.) Assume also that we know the function values $f(x,y)$ and the
first- and second-order derivatives, $f_x(x,y)$, $f_y(x,y)$ and
$f_{xy}(x,y)$ at all four corners. (Here subscript $x$ denotes
derivatives, $f_x = df/dx$.) The coefficients of the bicubic spline
are then defined such that that both the function values and the
derivatives match,
\begin{eqnarray}
f(x,y) &= p(x,y) \\
f_x(x,y) &= p_x(x,y)\\
f_y(x,y) &= p_y(x,y) \\ 
f_{xy}(x,y)&=p_{xy}(x,y),
\label{eq:spline_equations}
\end{eqnarray}
at all four corners. With four equations for each of four corners,
there is a total of 16 independent equations for the 16 independent
spline coefficients, $a_{ij}$, and the spline is therefore uniquely
defined. 

Writing out Equation \ref{eq:spline_equations} explicitly, we obtain
the following set of linear equations,
\begin{equation}
\begin{array}{cc}
\left.\begin{array}{ccccl}
f(0,0)& = &p(0,0) &= &a_{00}\\
f(1,0)& = &p(1,0) &= &a_{00} + a_{10} + a_{20} + a_{30}\\
f(0,1)& = &p(0,1) &= &a_{00} + a_{01} + a_{02} + a_{03}\\
f(1,1)& = &p(1,1) &= &\sum_{i=0}^3 \sum_{j=0}^3 a_{ij}
\end{array}\right\}
&  \textrm{Matching function values\quad\quad\quad\quad\quad\quad\quad} 
\end{array}.
\end{equation}

\begin{equation}
\begin{array}{cc}
\left.\begin{array}{ccccl}
f_x(0,0)& = &p_x(0,0) &= &a_{10}\\
f_x(1,0)& = &p_x(1,0) &= &a_{10} + 2a_{20} + 3a_{30}\\
f_x(0,1)& = &p_x(0,1) &= &a_{10} + a_{11} + a_{12} + a_{13}\\
f_x(1,1)& = &p_x(1,1) &= &\sum_{i=0}^3 \sum_{j=0}^3 i\,a_{ij}
\end{array}\right\}
&  \textrm{Matching first-order $x$-direction derivatives}
\end{array}
\end{equation}

\begin{equation}
\begin{array}{cc}
\left.\begin{array}{ccccl}
f_y(0,0)& = &p_y(0,0) &= &a_{01}\\
f_y(1,0)& = &p_y(1,0) &= &a_{01} + a_{11} + a_{21} + a_{31}\\
f_y(0,1)& = &p_y(0,1) &= &a_{01} + 2a_{02} + 3a_{03}\\
f_y(1,1)& = &p_y(1,1) &= &\sum_{i=0}^3 \sum_{j=0}^3 j\,a_{ij}
\end{array}\right\}
&  \textrm{Matching first-order $y$-direction derivatives}
\end{array}
\end{equation}

\begin{equation}
\begin{array}{cc}
\left.\begin{array}{ccccl}
f_{xy}(0,0)& = &p_{xy}(0,0) &= &a_{11}\\
f_{xy}(1,0)& = &p_{xy}(1,0) &= &a_{11}+2a_{21}+3a_{31}\\
f_{xy}(0,1)& = &p_{xy}(0,1) &= &a_{11}+2a_{12}+3a_{13}\\
f_{xy}(1,1)& = &p_{xy}(1,1) &= &\sum_{i=0}^3 \sum_{j=0}^3 ij\,a_{ij}
\end{array}\right\}
&  \textrm{Matching second-order $xy$ cross-derivatives}
\end{array}
\end{equation}

The remaining problem is then to estimate the first- and second-order
derivatives at all grid points, and several approaches may be used for
this. We adopt a spline based method for this step as well. 

First, we compute a standard one-dimensional natural spline along all
$x$ and $y$ coordinate lines, using standard methods. The result from
this operation is the set of all pure second-order derivatives,
$f_{xx}(x,y)$ and $f_{yy}(x,y)$, at each grid point. 

Because a one-dimensional spline is also a simple cubic polynomial,
and therefore only has four free coefficients, it is sufficient to
know the function values and second-order derivatives at all grid
corners to uniquely specify every coefficient. As a consequence of
this, the first-order derivative along the $x$-direction at grid point
$(x_i, y_j)$ is uniquely given by \citep{press:2002}
\begin{equation} 
f_{x}(x_i,y_j) = \frac{f(x_{i+1},y_j) - f(x_i,y_j)}{h_x} - \frac{1}{3}
h_x f_{xx}(x_i,y_j).
\end{equation}
Here $h_x = x_{i+1}-x_i$ is the $x$-direction grid cell size for the
current cell. An equivalent expression obviously holds for the
$y$-direction derivatives. 

Finally, to estimate the second-order cross-derivatives,
$f_{xy}(x,y)$, the above process is repeated such that $y$-derivatives
are computed from $f_x(x,y)$ splines for all $y$-direction coordinate
lines. Thus, all required derivatives may be obtained by performing $m
+ 2n$ one-dimensional spline computations, where $m$ is the number of
grid cells in $x$-direction, and $n$ is the number of grid cells in
$y$-direction.


\begin{thebibliography}{}

\bibitem[Ackerman et al.(2007)]{ackerman:2007} Ackerman, L., Carroll, 
S.~M., \& Wise, M.~B.\ 2007, \prd, 75, 083502 

\bibitem[Bennett et al.(2003)]{bennett:2003} Bennett, C. L., et al.\
  2003, \apjs, 148, 1

\bibitem[Dunkley et al.(2009)]{dunkley:2009} Dunkley, J., et al.\ 
2009, \apjs, 180, 306 

\bibitem[Eriksen et al.(2007)]{eriksen:2007} Eriksen, H.~K., et al.\ 
2007, \apj, 656, 641 

\bibitem[G\'orski et al.(2005)]{gorski:2005} G\'orski, K. M., Hivon, E.,
  Banday, A. J.,Wandelt, B. D., Hansen, F. K., Reinecke, M.,
  Bartelman, M. 2005, \apj, 622, 759

\bibitem[Green \& Silverman(1994)]{green:1994} Green, P. J.,
  \& Silverman, B. W.\ 1994, Non-Parametric Regression and Generalized
  Linear Models, Chapman and Hall, 1994

\bibitem[Groeneboom 
\& Eriksen(2009)]{groeneboom:2009a} Groeneboom, N.~E., \& Eriksen, H.~K.\ 2009, \apj, 690, 1807 

\bibitem[Groeneboom et al.(2009)]{groeneboom:2009b} Groeneboom, N.~E., 
Eriksen, H.~K., Gorski, K., Huey, G., Jewell, J., 
\& Wandelt, B.\ 2009, arXiv:0904.2554 

\bibitem[Hill et al.(2009)]{hill:2009} Hill, R.~S., et al.\ 2009, 
\apjs, 180, 246 

\bibitem[Hinshaw et al.(2003)]{hinshaw:2003} Hinshaw, G., et al.\ 
2003, \apjs, 148, 63 

\bibitem[Hinshaw et al.(2007)]{hinshaw:2007} Hinshaw, G., et al.\ 
2007, \apjs, 170, 288 

\bibitem[Hinshaw et al.(2009)]{hinshaw:2009} Hinshaw, G., et al.\ 
2009, \apjs, 180, 225 

\bibitem[Hivon et al.(2002)]{hivon:2002} Hivon, E., G{\'o}rski, 
K.~M., Netterfield, C.~B., Crill, B.~P., Prunet, S., 
\& Hansen, F.\ 2002, \apj, 567, 2 

\bibitem[Huffenberger et al.(2006)]{huffenberger:2006} Huffenberger, 
K.~M., Eriksen, H.~K., \& Hansen, F.~K.\ 2006, \apjl, 651, L81 

\bibitem[Jarosik et al.(2007)]{jarosik:2007} Jarosik, N., et al.\ 
2007, \apjs, 170, 263 

\bibitem[Komatsu et al.(2009)]{komatsu:2009} Komatsu, E., et al.\ 
2009, \apjs, 180, 330 

\bibitem[Lewis \& Bridle(2002)]{lewis:2002} Lewis, A., \& Bridle, S.\ 2002, \prd, 66, 103511 


\bibitem[Nolta et al.(2009)]{nolta:2009} Nolta, M.~R., et al.\ 
2009, \apjs, 180, 296 

\bibitem[Page et al.(2003)]{page:2003} Page, L., et al.\ 2003, 
\apjs, 148, 39 

\bibitem[Press(2002)]{press:2002} Press, W.~H.\ 2002, ``Numerical 
recipes in C++ : the art of scientific computing'', Cambridge
University Press, ~ISBN: 0521750334

\bibitem[Spergel et al.(2007)]{spergel:2007} Spergel, D.~N., et al.\ 
2007, \apjs, 170, 377 

\bibitem[Wandelt \& G{\'o}rski(2001)]{wandelt:2001} Wandelt, B.~D., \& G{\'o}rski, K.~M.\ 2001, \prd, 63, 123002 

\bibitem[Wright et al.(1996)]{wright:1996} Wright, E.~L., Hinshaw, 
G., \& Bennett, C.~L.\ 1996, \apjl, 458, L53 

\end{thebibliography}
\end{document}